\documentclass[a4paper, 9pt, twocolumn]{article}
\usepackage[pass]{geometry}
\usepackage[utf8]{inputenc}
\usepackage[english]{babel}
\usepackage{csquotes}
\usepackage[cm]{fullpage}
\usepackage{graphicx}
\usepackage{authblk}
\usepackage[per-mode = symbol, separate-uncertainty = true]{siunitx}
\usepackage[
style = nature,
backend = biber,
citestyle = nature,
url = false
]{biblatex}
\bibliography{references}

\title{Magnetic resonance imaging with optical preamplification and detection}
\author[1]{A. Simonsen}
\author[2]{J. D. Sanchez}
\author[1,3]{S. A. Saarinen}
\author[2]{J. H. Ardenkj{\ae}r-Larsen}
\author[1,3]{A. Schliesser}
\author[1]{E. S. Polzik}
\affil[1]{Niels Bohr Institute, University of Copenhagen}
\affil[2]{Center for Hyperpolarization In Magnetic Resonance (HYPERMAG), Technical University of Denmark}
\affil[3]{Center for Hybrid Quantum Networks (Hy-Q), University of Copenhagen}
\date{}

\begin{document}
\maketitle


\begin{abstract}
    Magnetic resonance (MR) imaging relies on conventional electronics that is increasingly challenged by the push for stronger magnetic fields and higher channel count. These problems can be avoided by utilizing optical technologies. As a replacement for the standard low-noise preamplifier, we have implemented a new transduction principle that upconverts an MR signal to the optical domain and imaged a phantom in a clinical \SI{3}{\tesla} scanner with signal-to-noise comparable to classical induction detection.
\end{abstract}

\section{Main}
\begin{figure}
    \centering
    \includegraphics{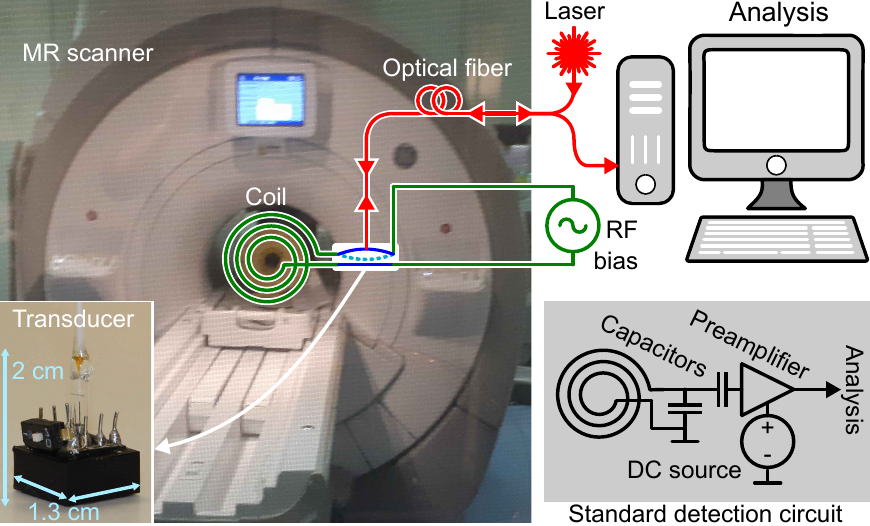}
    \caption{\emph{Simplified schematic of electro-mechano-optical transduction.} $\,$
    The main photograph shows the MR scanner with the vacuum chamber inside. In the sketch of the scheme, an external frequency bias powers the detection scheme to light and optical fibers routes light from laser to transducer (inset lower left), and from transducer to analysis. The coil can be any standard RF coil used for MR detection. In the standard MR detection scheme (lower right), the RF coil is matched to a low-noise preamplifier. A DC voltage source powers the amplifier and electrical cables carry the amplified signal out of the scanner for further processing.
    } 
    \label{fig:combi}
\end{figure}

Magnetic resonance (MR) imaging is a well-established and non-invasive tool for routine clinical diagnostics and basic research. Its utility would benefit from enhanced signal-to-noise and acquisition speed---the key performance parameters for MR imaging---and is the focus of considerable research efforts.
Two trending developments are higher magnetic fields and more detection coils in arrays.
Yet most technical innovation relies on the same principle of matching a detection coil to a low-noise preamplifier in the receiver chain of the scanner \cite{Edelstein1986} similar to the diagram in Fig. \ref{fig:combi} (lower right), a common practice that testifies to the extraordinary performance of conventional electronics obtained through decades of optimization.
However, the electronics is increasingly challenged by the ever-increasing magnetic field strengths in MR imaging \cite{Sobol2012}. And with the push for more coils, the standard approach adds problems like cross-talk and electrical interference associated with the individual preamplifiers and cables that each coil needs, not to mention that dense arrays demands a miniaturization of the corresponding circuits \cite{Wiggins2009}.
These issues may be alleviated by leveraging optical communication and sensing. In particular, optical fibers are insensitive to high magnetic fields, immune to electrical interference, and considered MR safe \cite{Taffoni2013}. Converting the MR signal into an optical modulation has been done \cite{Memis2008}, and is even commercially available (Phillips dStream), but those approaches still use the standard electrical preamplifier first and then convert the amplified signal.
In contrast, this work replaces the conventional electrical preamplifier with a transducer \cite{Bagci2014, Takeda2017, Simonsen2019} that up-converts the MR signal onto an optical carrier. The signal can then be analyzed at the other end of a long and low-loss optical fiber which moves the receiver electronics out and away from the coils and high magnetic field in the scanner. The idea is sketched in Fig. \ref{fig:combi} where a signal induced in an MR coil is combined with a radio-frequency (RF) bias, and together they vibrate a micro-mechanical element which, in turn, modulates the amplitude of reflected laser light.
The transduction scheme can have a low noise-temperature \cite{Bagci2014}, can target any signal frequency, is compatible with telecom optical wavelengths, and has been successfully used to detect a nuclear magnetic resonance signal \cite{Takeda2017, Tominaga2018}. And with proper circuit design and protective steps, we here show for the first time that it can be integrated into a commercial, clinical MR scanner to acquire and reconstruct a complete MR image.

Fig. \ref{fig:combi} shows a simplified version of our detection setup. A detailed diagram is in the supplementary information. Inside an MR imaging scanner (\SI{3}{\tesla} GE MR750), a typical MR RF coil formed a resonant circuit together with a mechanically compliant capacitor---the transducer. The circuit was tuned to resonate at the Larmor precession frequency of the nuclear spins aligned along a strong magnetic field. In our case, the setup specifically involved $^{13} C$ nuclear spins and a \SI{3}{\tesla} magnetic field, i.e. a precession of \SI{32}{\mega\hertz}.
The scanner applied an RF pulse to excite the nuclear precession---the transmit pulse. Furthermore, it applied gradient magnetic fields that enable selective excitation of atoms and encodes their phase through spatially-dependent time evolution. As the spins precess, they create an oscillating magnetic flux through the coil thus inducing a voltage in the RF coil. That voltage is usually electronically amplified and processed to generate the MR image.
Our new development consists of replacing the standard preamplifier with an optomechanical transducer that converts the MR signal to an amplitude modulation of laser light. The transducer was connected directly in parallel to the coil and consisted of a freely suspended membrane---equivalent to a tiny drum-skin---that simultaneously constitutes one side of a capacitor and one mirror in an optical cavity. Charges on the capacitor exert a force on the membrane and cause it to move, and this motion then changes the capacitance and thus also the charges that affects the motion. Hence, the mechanical and electrical resonance are parametrically coupled \cite{Dougherty1996}. As the MR signal induces charges on the capacitor, the resulting motion proportionally changes the reflection of light from the optical cavity. This transduction is most sensitive when the membrane is driven near resonance, which a signal at any frequency can do in unison with an RF bias if the beatnote between the signal and the bias is at the membrane's resonance---see \cite{Zeuthen2017a} for the detailed theory. In our case, that meant the bias frequency had to be the difference between the Larmor precession and the membrane resonance, \SI{32}{\mega\hertz} and \SI{1.4}{\mega\hertz} respectively. Note that because the bias and signal are non-degenerate, interference and cross-talk between them can be removed through filtering and the mechanical response helps with that.
After the transduction, we downconverted the detected signal from optical modulation to the RF domain simply by measuring the reflected optical power with a detector. Although this output was frequency-shifted to the mechanical frequency, it corresponded to the voltage induced in the coil by the MR sequence and was consequently sufficient to reconstruct the MR image through post-processing. However, we chose, out of convenience, to leverage the scanner's dedicated imaging software and fed the transduced signal back to the scanner's receiver channel. That required the signal to be frequency-shifted back to the signal frequency expected by the scanner, which we did by mixing the transducer signal with the RF bias (see the methods section and figure in supplementary information).

The transducer's membrane was circular and made out of alumina and aluminum under tensile stress attached on top of a partially reflective mirror. See the methods section and \cite{Simonsen2019} for details. Because aluminum reflects, the membrane and mirror formed an optical cavity with a length determined by the cleanroom fabrication procedure. We designed the stack such that the fixed wavelength of our laser was about one cavity linewidth (half width at half max) away from the resonance of the optical cavity. Laser light was routed to the sample by an optical network and coupled directly from fiber into the cavity through focusing optics. The same fiber also collected the reflection, and the optical network then routed the signal to a detector.
To electrically connect the transducer and the circuit, we mounted and wirebonded the membrane chips to an 8-pin integrated circuit socket. Because the transducer needs vacuum to operate, we placed the assembly inside a vacuum chamber made of glass-fiber that could hold a pressure below \SI{1e-3}{\milli\bar} throughout the measurement series---see the method section. The chip and the circuit were then connected with a short RF cable via a vacuum feedthrough.
We protected the transducer from electro-static discharges with a switch (see Fig. \ref{fig:combi}, lower left) that could short-circuit the capacitor pins. This was particularly important whenever we connected the transducer to circuitry or transported it.
In addition, we took several protective steps beyond \cite{Bagci2014, Simonsen2019, Haghighi2018, Takeda2017, Tominaga2018} to prevent the membrane from collapsing during the transmit pulse: first, two crossed diodes were added in parallel to the transducer to short it if the induced voltage exceed their threshold. Such protection is a common way to protect standard preamplifiers, but did not sufficiently protect the transducer on its own. In addition, it unfortunately limits the RF bias that can be applied, but we could anyway get satisfactory transduction performance. Second, the detection resonance was detuned to minimize the voltage and current induced in the coil by the transmit pulse. This is another standard technique in MR imaging \cite{Edelstein1986} and the scanner supplied the trigger signal that controlled the detuning. For this purpose, the coil loop had a segmenting capacitor and additional components that together created a tank circuit when the trigger activated a PIN diode's forward-voltage threshold. The full circuit diagram is described in the methods section and shown in supplementary information. Third, the RF bias was switched off by pulse-modulating the bias amplitude with the same trigger that detuned the circuit. See the full setup diagram in supplementary information. Note that the RF bias and spin-flip pulse overlap a little because the long (\SI{8}{\meter}) RF cables to and from the scanner add a time delay (\SI{\sim 100}{\nano\second}), but this was not a problem with everything else in place.

\begin{figure*}
    \centering
    \includegraphics[scale=1.0]{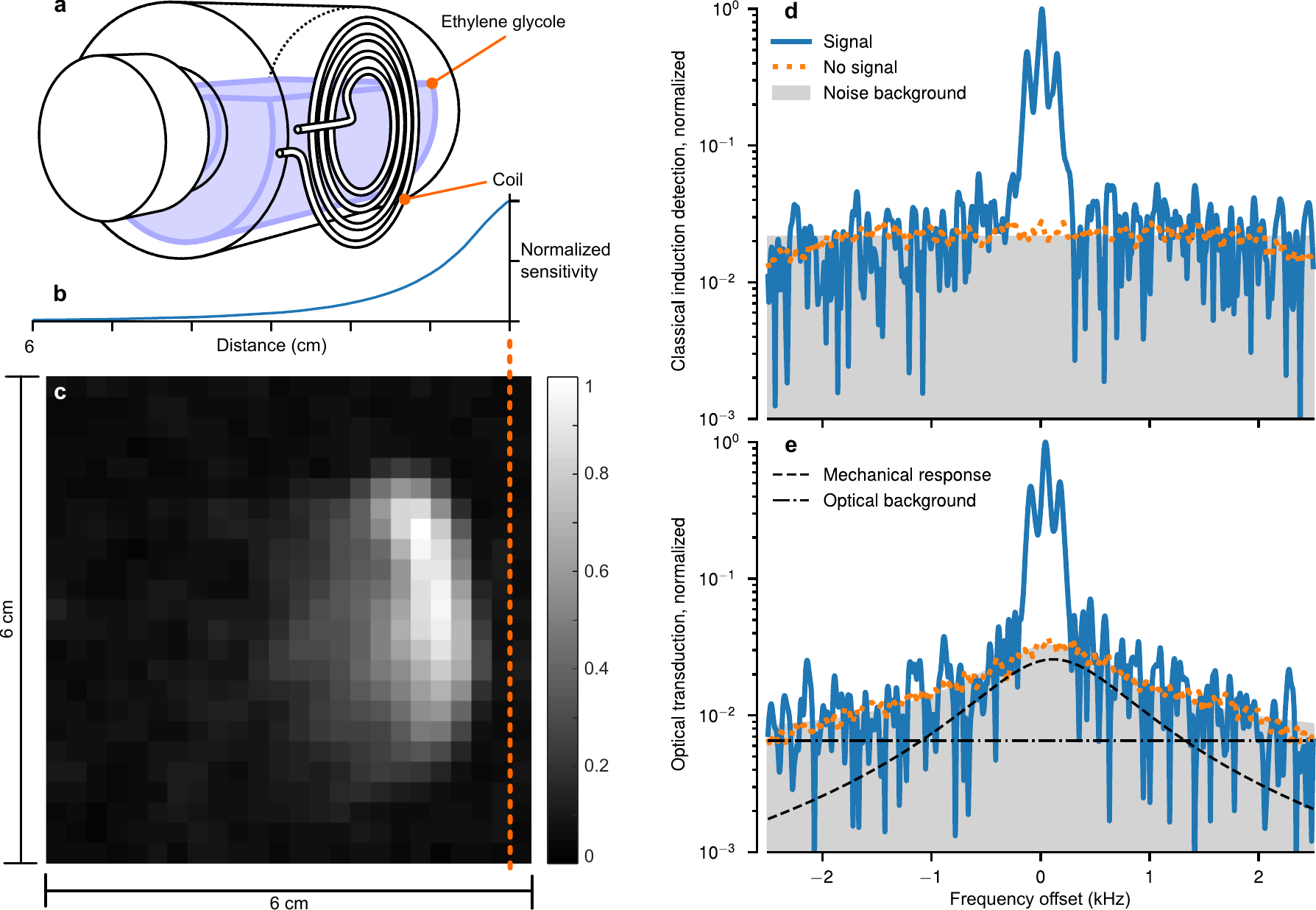}
    \caption{\emph{MR image data obtained with the electro-mechano-optical transduction} $\,$
    \textbf{a}, Sketch of the coil and phantom setup. The image plane intersects the center of the coil and is parallel to the bottle circumference.
    \textbf{b}, Simulated sensitivity versus distance to the coil.
    \textbf{c}, Surface plot showing the normalized peak amplitude of the detected signal. The dashed line indicates where the plane of the coil intersects the image.
    \textbf{d}, Spectrum of the MR signal measured with classical induction detection. \textbf{e}, Same spectrum measured with the transducer. Both spectra shows data with and without an induced MR signal, normalized to the signal-peak. With a signal, the spectrum shows the highest peak among the image voxels; without, the spectrum shows the average over the entire image and is equivalent to the noise background of detection.} 
    \label{fig:results}
\end{figure*}

In its first implementation, the transduction scheme had problems with excess noise which we addressed in the following ways:
first, the RF bias was filtered twice to reduce its sideband noise at the detection frequency. Once through an external filter and then through a filter built into the detection circuit. Such sideband noise can limit sensitivity \cite{Takeda2017, Tominaga2018}.
Second, there was considerable added noise when flexible cables connected the setup inside the scanner room with the setup outside. We believe the poor shielding of the cables allowed ambient noise to leak into the setup. Using instead semi-rigid cables with better shielding reduced the noise significantly. Third, electrical noise at the membrane-frequency could drive the mechanical motion because our sample had trapped charges \cite{Simonsen2019}. We cancelled these out with a DC bias on the transducer which gave significant reduction in the overall transducer noise. With those noise-eliminating steps in place, we obtained the data shown in Fig. \ref{fig:results} with the optomechanical transduction.


We imaged a phantom (Fig. \ref{fig:results}a) consisting of a bottle filled with ethylene glycol (purity \SI{99.8}{\percent}; natural abundance, \SI{1.1}{\percent}, $^{13} C$; triplet with $J_{CH}$ of \SI{142}{\hertz}). The recorded image (Fig. \ref{fig:results}c) shows the spatial density of $^{13} C$ atoms in a cross-section through the bottle. As we expected, the image does not have a uniform signal-to-noise throughout the volume because we used a surface coil where the detected signal decays with the increased distance to the coil (Fig. \ref{fig:results}b). This effect is not corrected in the image processing.
To each image voxel corresponds a spectrum obtained from the spin-precession (Fig. \ref{fig:results}d). For ethylene glycol, that spectrum has a characteristic triplet as shown in the reference measurement (Fig. \ref{fig:results}d), obtained with a commercial coil but still using the same phantom and scanner. And the spectrum obtained with the transducer (Fig. \ref{fig:results}e) also shows the same triplet. Additionally, the two detection schemes have different background noise; for the standard electronic amplifier the noise is flat in a broad window, but for the transducer it has a distinctive narrow Lorentzian lineshape plus an offset (Fig. \ref{fig:results}e). 
We believe the Lorentzian feature comes from the membrane's spectral response because the Lorentzian peak frequency and linewidth changed with the bias power as we expect from the electromechanical interaction \cite{Bagci2014, Zeuthen2017a}. Additionally, moving the bias frequency also shifted the detected signal with respect to the noise peak and scaled its amplitude following the peak shape. This is an expected behaviour if the MR signal drives the membrane's motion and therefore further corroborates that the spectral shape of the noise is related to the mechanical motion. Furthermore, the transducer's flat noise background scales with optical power which suggests that it originates from amplitude noise of the laser, most likely due to shot-noise of light. Finally, we have not found any other spectral feature in the circuit response, nor any spurious noise in the setup, that is consistent with the observed Lorentzian peak.

In summary, we have implemented MR imaging with direct optical detection and amplification of the MR induction signal thus bringing the technical benefits of optical signal processing to the receiver chain of an MR scanner. To the best of our knowledge, this work represents the first successful implementation of direct electro-mechano-optical transduction in an MR system.
We clearly see the expected image and spectrum for our phantom and coil, and a noise-background compatible with the mechanical motion's Lorentzian linewidth plus an offset from optical noise. The transducer's spectrum shows a signal-to-noise comparable with a commercial system, although still not as high, and a transduction bandwidth that is narrow compared to standard preamplifiers.
However, both features can be improved greatly with straightforward improvements such as increasing the RF bias amplitude, reducing the capacitance in parallel with the transducer, and especially by decreasing the membrane-capacitor gap \cite{Takeda2017}. The noise can be reduced by cooling the system, increasing the mechanical quality factor, and using a lighter membrane \cite{Tominaga2018}. Bandwidth can be increased by using multiple mechanical modes \cite{Haghighi2018}. Improvements of the optical cavity will also benefit performance. Incidentally, our sample had a suboptimal cavity length for the specific bias voltage, so we can expect to reach better sensitivity using the present platform but with some fabrication optimization.
Note that we deliberately reduced signal-to-noise slightly by upconverting the detected signal in order to send it back to the scanner. This step was not necessary, but convenient.
Our future work will aim to address the trapped charges in transducer fabrication in order to eliminate the DC bias, and to address the transducer's need for vacuum with cleanroom packaging. The required vacuum should be achievable with available techniques \cite{Rushton2014}. Furthermore, while the RF bias is necessary to power the transduction scheme, supplying it with cables is not a necessity. The bias could instead be delivered wirelessly \cite{Menke2017}, potentially delivered by the same coil that generates the transmit pulses, and collected by the RF coil. Such a scheme would eliminate all electrical connections to the RF coil, leaving the optical fiber as the only physical connection to the circuitry. This would enable a great increase in the density of elements in MR array coils, avoiding all-together the performance and safety issues that stems from the large number of electrical connections present in MR arrays based on current technology.

\printbibliography

\newpage
\section{Methods}

\subsection{Transducer fabrication}
We fabricated the transducer in a cleanroom using standard techniques---see \cite{Simonsen2019} for a step-by-step account. The starting substrate was a fused silica wafer, \SI{100}{\milli\meter} in diameter, with a partially reflective dielectric mirror on top. Alumina layers both protected the mirror and supported the membrane while aluminum defined the top and bottom electrodes of the membrane-capacitor. The membrane consisted of \SI{\sim 70}{\nano\meter} alumina and \SI{\sim 90}{\nano\meter} aluminum. We added tensile stress to the aluminum by annealing the wafers before releasing the membrane. Alumina got its tensile stress during deposition. A sacrificial silicon nitride layer was between the electrodes and it had a thickness designed to the desired cavity length. The layer also determines the gap between the capacitor electrodes. At the end of the fabrication, the layer is etched away to release the membrane.

The bottom electrode contained a hole, aligned to the center of the membrane, where light could pass through. Laser light was coupled into the cavity through this hole, using a gradient refractive-index lens attached to the backside of the transducer chip. We achieved optical alignment by first centering the membrane with respect to a silicon chip attached to the backside of the transducer chip. The silicon die had a large, square, and centered hole that guided the lens and ensured alignment to the cavity. Light was supplied through a fiber-pigtail terminated in a glass ferrule, with the ferrule and lens aligned by a glass tube that snugly fitted both. By changing the distance between ferrule and lens, we maximized the light delivered by the fiber, focused through the lens, reflected from the sample, and collected back into the input fiber. Finally, all components were fixed permanently with glue.

\subsection{Circuit}
The full circuit used to detect the imaging signal is shown in the supplementary information. The RF coil was a flat spiral coil with four windings and an outer diameter of \SI{50}{\milli\meter}, wound with a \SI{1.6}{\milli\meter} diameter silver wire. It had an inductance estimated to be \SI{490}{\nano\henry} from simulation. A segmenting capacitor was inserted \cite{Rispoli2016} which formed a trap circuit together with parallel circuitry. The trap detuned the detection coil when activated by a PIN diode, i.e. when the diode's forward-bias threshold was exceeded by transmit trigger voltage.
The main circuit board also included a bandpass filter at the RF bias frequency. Its purpose was to filter the RF bias' sideband noise at the MR signal frequency and prevent the RF bias from loading the detection resonance. Notably, the filter was designed to allow a DC offset on the bias to pass through.

\subsection{MR setup}
Supplementary information shows the full setup diagram explained here.
We mounted the RF coil and detection circuit right outside a cryostat made from glass-fiber. Although the cryostat can cool the transducer and circuit, we only used it as a vacuum chamber. It could maintain vacuum below \SI{1e-3}{\milli\bar} up to five hours after disconnecting from its pump because it contained two molecular sieves (activated charcoal and sodium aluminum silicate), both cooled to liquid nitrogen temperature (\SI{77}{\kelvin}). When mounting the transducer in the cryostat, we tried to align the membrane perpendicular to the main magnetic field. Without this alignment, the mechanical linewidth broadens, likely due to the Lorentz force on the charges in the membrane.
Laser light came to the transducer through a custom fiber-feedthrough \cite{Abraham1998} for the cryostat, an optical 90/10 splitter, and a long single-mode fiber connected directly to the fiber-coupled \SI{1064}{\nano\meter} laser outside the scanner room. The splitter only routed \SI{10}{\percent} of input light to the chip and dumped the remaining power. Light reflected from the transducer's cavity went back into the same splitter which then distributed \SI{90}{\percent} into another long single-mode fiber that lead back outside and connected to a custom-built detector.

Inside the scanner room, we connected the trigger signal directly to the detuning trap on the circuit and also sent the trigger outside to modulate the RF bias. That same cable also carried the transduced signal back into the scanner. Two bias-tees handled the routing by frequency discriminating the low-frequency trigger and the high-frequency signal. A separate cable carried the RF bias.
Outside the scanner room, the RF bias had its pulse-modulation input connected to the trigger and its output split 50/50. One part became the bias drive and we amplified it before sending it to the detection circuit through and external filter. The other part became the local oscillator in a mixer that frequency-shift the optically detected signal back to the MR signal frequency. The output of this mixer was the input of the scanner's receiver channel. We added a DC bias onto the RF bias drive with another bias-tee, right after the first bandpass filter for the RF bias and before the long cable going into the scanner room.

\newpage

\section*{Supplementary information}


\begin{figure}[!h]
    \centering
    \includegraphics{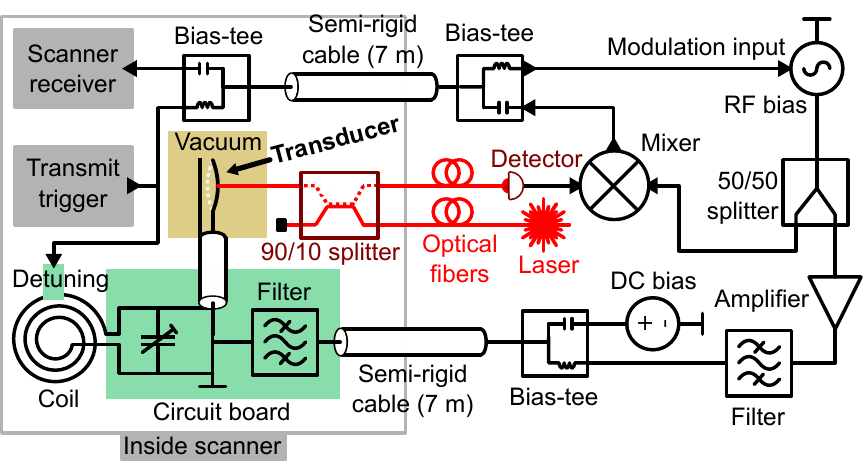}
    \caption{\emph{Detail setup schematic.}
    } 
\end{figure}

\begin{figure}[!h]
    \centering
    \includegraphics{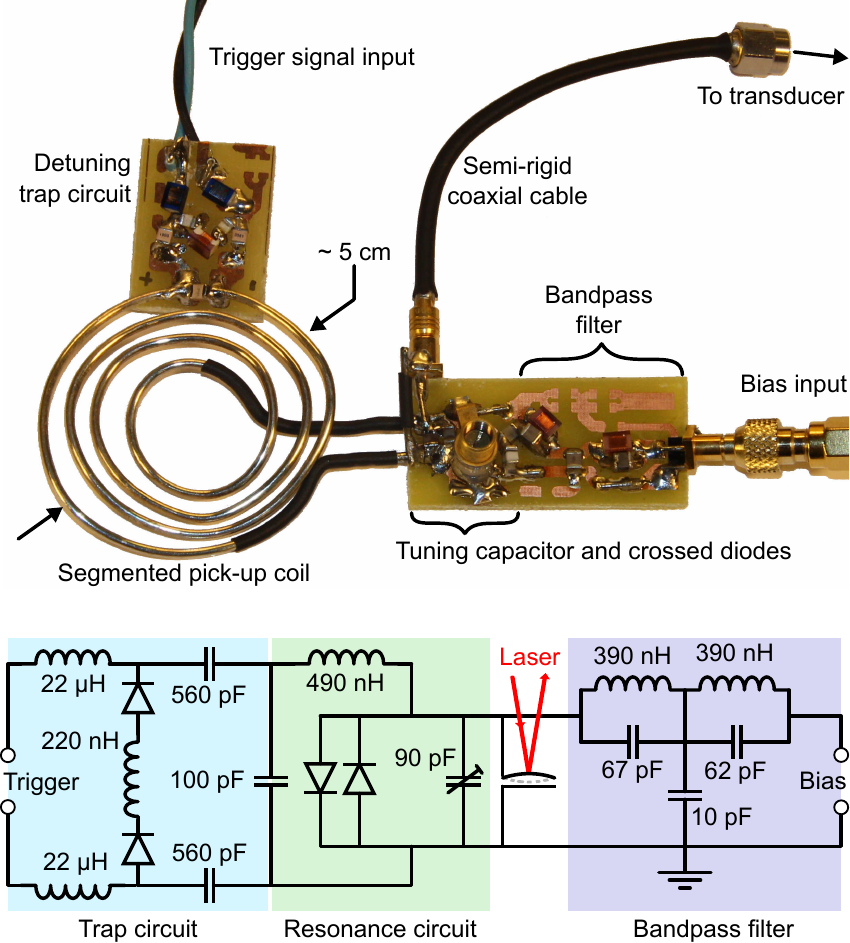}
    \caption{\emph{Photo (top) and detailed diagram (bottom) of the detection circuit}
    }
\end{figure}

\end{document}